\documentclass[cits]{PoS}  

\usepackage{wrapfig}
\usepackage[utf8]{inputenc}
\usepackage{microtype}
\usepackage{amsmath}

\DeclareMathOperator{\sort}{sorted}

\DeclareMathOperator{\Arg}{Arg}

\newcommand{\T}{\dag}
\newcommand{\X}{\mathcal{X}}
\newcommand{\n}{\mathcal{N}}
\newcommand{\G}{\mathcal{G}}

\newcommand{\F}{\mathcal{F}}
\newcommand{\U}{\mathcal{U}}

\newcommand{\eX}{\mathbb{X}}
\newcommand{\eN}{\mathbb{N}}

\title{Ensemble Quasi-Newton HMC}

\ShortTitle{Ensemble Quasi-Newton HMC}

\author{\speaker{Xiao-Yong Jin} and James C. Osborn\\
  Computational Science Division\\
  Argonne National Laboratory\\
  9700 S. Cass Ave.\\
  Lemont, IL 60439, USA\\
  E-mail: \email{xjin@anl.gov}, \email{osborn@alcf.anl.gov}}

\abstract{We present a modification of the Hybrid Monte Carlo algorithm for
tackling the critical slowing down of generating Markov chains of
lattice gauge configurations towards the continuum limit.  We propose
a new method to exchange information within an ensemble of Markov
chains, and use it to construct an approximate inverse Hessian
matrix of the action inspired from quasi-Newton algorithms for
optimization.  The kinetic term of the molecular dynamics evolution
includes the approximate Hessian for long distance couplings among
the momenta.  We show the result of applying the new algorithm
to the $U(1)$ gauge theory in two dimensions, and discuss our future
plans.}

\FullConference{The 36th Annual International Symposium on Lattice Field Theory - LATTICE2018\\
		22-28 July, 2018\\
		Michigan State University, East Lansing, Michigan, USA.}

\begin{document}

\vspace{-0.4em}
\section{\label{sec:introduction}Introduction}
\vspace{-0.5em}

In generating a Markov chain, we aim at speeding up Monte Carlo
simulations, making proposal configurations far from the
current configuration in phase space, with relative low cost.
Molecular Dynamics (MD) evolution in fictitious time using
random momenta naturally extends and mitigates the Langevin-like random walk
behavior.
This Hybrid Monte Carlo (HMC)
algorithm~\cite{Duane:1987de} works well in high dimensional
systems, such as lattice QCD.
Approaching the continuum limit of the lattice theory,
some physical modes in MD slows down exponentially,
leading to research in
Fourier acceleration~\cite{Batrouni:1985jn,Duane:1986fy,Duane:1988vr}
as a possible remedy.
The analogous Riemannian manifold HMC~\cite{RSSB:RSSB765}
claims success for some probability density functions
in guiding the MD evolution through the phase space.
Recent efforts~\cite{Cossu:2017eys,Christ:2018net,Zhao:2018jas} surge
in analyzing and applying similar techniques to lattice QCD.
We focus on employing, as the acceleration kernel, a
numerically cheaper approximation of the Hessian matrix from the
Limited-memory Broyden-Fletcher-Goldfarb-Shanno
algorithm~\cite{Fletcher:1970,Nocedal:1980} (L-BFGS), a common
quasi-Newton optimization method.  This approximation applies to
not only the gauge fields but also the pseudo-fermion fields.
The HMC, nevertheless, requires changes to adopt such
approximation that uses information from multiple configurations.

In this paper we present the
ensemble quasi-Newton HMC (QNHMC) method,
discuss the characteristics of the method on two-dimensional $U(1)$
lattice gauge theory,
and show preliminary results of its effect on the autocorrelation of
the average plaquette value and topological charge.

\vspace{-0.4em}
\section{\label{sec:markov-chain}Markov chain for assisted MD evolution}
\vspace{-0.5em}

The MD evolution in the heart of the HMC algorithm follows from Hamiltonian dynamics,
\begin{equation}
\label{eq:md}
	H(x,p)=S(x)+\frac{1}{2}p^{\T}G^{-1}p,\quad
	\dot{x}=G^{-1}p, \text{ }
	\dot{p}=-\nabla S,
\end{equation}
where $S$ is the action, $p$ the
fictitious momenta, and $G$ a fixed MD mass matrix.
A symplectic and reversible discrete integrator advances
the state of the Markov chain from $(x,p)$ to $(x',p')$
over a fictitious time period,
$\tau$, the trajectory length.
Using the Hamiltonian as the negative log probability of the
enlarged phase space including $x$ and $p$, the correctness of
the HMC demands a positive definite $G$.

The choice of $G$ affects the performance of HMC.
For a general action, a MD mass matrix containing the local
information of the Riemann curvature can bring considerable
speedups~\cite{RSSB:RSSB765} in the efficiency of Markov Chain
Monte Carlo methods.  The article recommends the Fisher
information matrix as $G$.
Fourier acceleration~\cite{Duane:1986fy,Duane:1988vr} suggests
the field Laplacian operator as $G$.
We are interested in using a fixed $G$ during one MD trajectory,
for its simplicity and efficiency.  Any explicit symplectic
reversible integrator for equation~\eqref{eq:md} would still be
applicable.  However, this also means that $G$ cannot depend on
any configurations from this one whole MD trajectory.

In general we want a proposal of $(x',p')$ for the next state of
the Markov chain from $(x,p)$ following a symplectic and
reversible discretization of the MD evolution~\eqref{eq:md} where
$G^{-1}$ comes from our choice of a function, $\G^{-1}$,
\begin{equation}
  \label{eq:md-assisted-G}
  G^{-1} = \G^{-1}\big(\{\X_i\}_{i=0}^{\n-1}\big).
\end{equation}
whose argument is a list of $\n$ configurations $\X_i$,
which are all different from $x$, $x'$, or any
configuration along the discretized path of this particular MD
evolution.
Assuming a particular choice of $G^{-1}$ could help through
out the Markov chain generation, we can fix $\G^{-1}$ and
$\{\X_i\}_{i=0}^{\n-1}$,
optimally picked from available configurations,
and the HMC procedure remains the same except for
the additional mass matrix $G$.

In this paper, we focus on building $G^{-1}$ that is fixed
during one MD evolution but changes
after each trajectory.
We use $\{\X_i\}_{i=0}^{\n-1}$ from
a set of parallel streams of Markov chains.
We update
one of the streams using information from the others, suggested in
references~\cite{NIPS2011_4464,Matthews:2016aa}.  We can have
multiple ways to generate such Markov chains, and obtain $\G$ of
$\X$ from neighboring streams.  The following is the base case
with provable reversibility.  We use an arbitrary information
exchange kernel $\F$ to generalize the ensemble assisted Markov
chain.

Let $\eN$ be the number of coupled parallel streams, each labeled
$\eX_j$, for $j=0$ to $\eN-1$.  Let $\F$ be a function on a
\emph{unordered} set of configurations, $\{\X_i\}_{i=0}^{\n-1}$, with
$\n=\eN-1$.  Let $\U$ be a symplectic and reversible mapping that
generates the next state of one Markov chain, from $\eX_j$ to
$\eX'_j = \U\left(\F\big(\{\X_i\}_{i=0}^{\n-1}\big)\right)\eX_j$.
Given a fixed set of $\{\X_i\}_{i=0}^{\n-1}$, $\U$ depends on the
value of $\F$, and satisfies the detailed balance,
$\pi(\eX_j)P(\eX_j|\eX'_j) = \pi(\eX'_j)P(\eX'_j|\eX_j)$, where
$\pi$ is probability density we want to simulate and $P$ the
transition probability.  We give the definition of
$\{\X_i\}_{i=0}^{\n-1}$ within the steps of the ensemble assisted
Markov chain described in the following.
\begin{enumerate}
\item When updating $\eX_k$ for each $k$ from $0$ to $\eN-1$:
  \begin{enumerate}
  \item Setting the list, $\{\X_i\}_{i=0}^{\n-1}$ from the list
    $\{\eX_j\}_{j\ne k}$, with $\n=\eN-1$.
  \item Evolve $\eX_k$ according to
    $\U\left(\F\big(\{\X_i\}_{i=0}^{\n-1}\big)\right)$.
  \end{enumerate}
\item After generating one trajectory for each $\eN$ parallel
  streams, Set the new $\{\eX'_j\}$ to the reversed sequence of
  it, $\eX'_j\leftarrow\eX'_{\eN-1-j}$.  This is for the purpose of
  reversibility.
\end{enumerate}

\begin{wrapfigure}{r}{0.4\textwidth}
  \centering
  \includegraphics[width=0.44\textwidth]{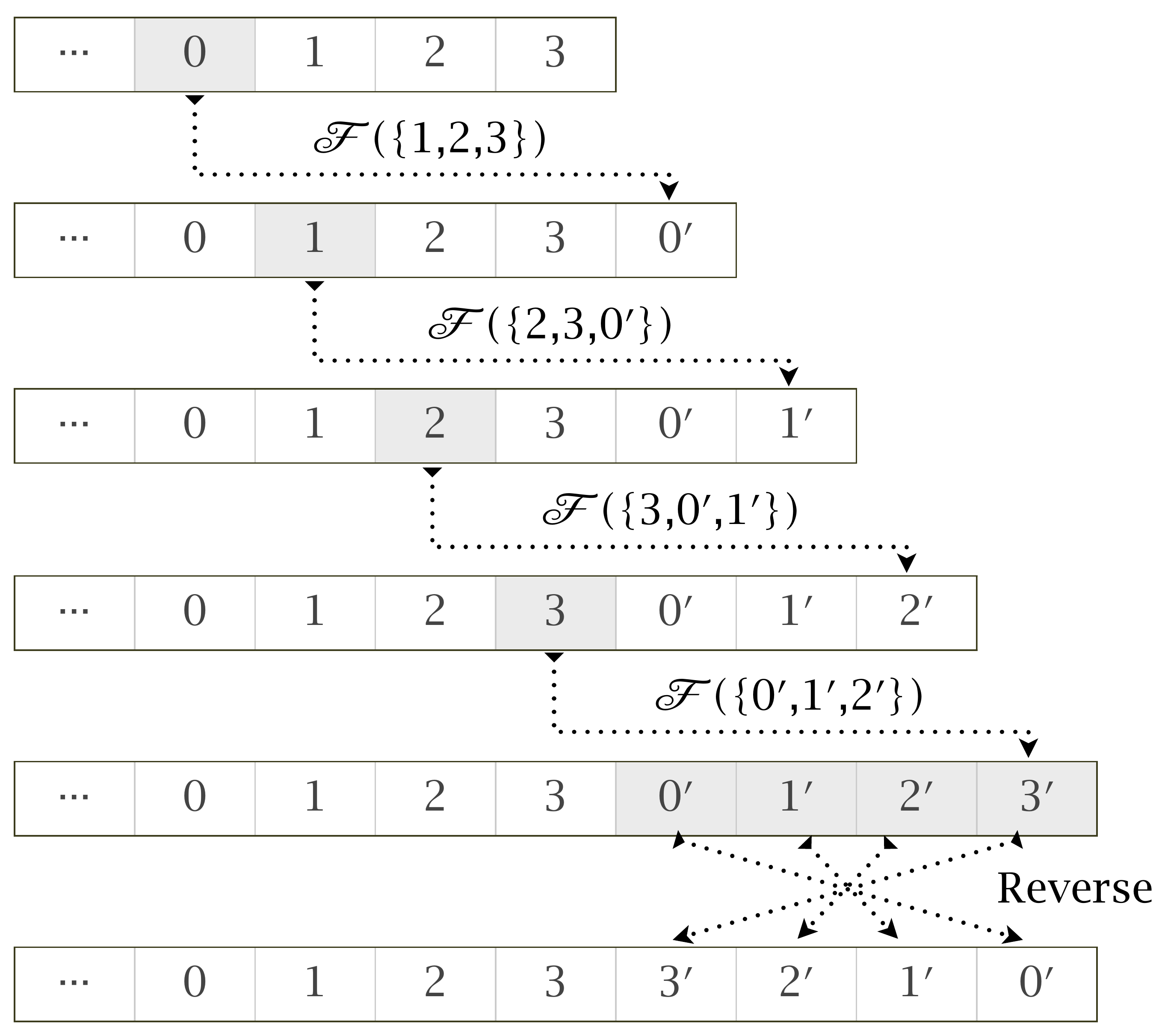}
  \caption{\label{ensHMC}One update for $\eN=4$.}
\end{wrapfigure}

Figure~\ref{ensHMC} illustrates an example of one update with
$\eN=4$ streams.  The current configurations from these four
streams are labeled with $0$, $1$, $2$, and $3$.  We update those
successively, each time with information obtained from the
function $\F$ applied to configurations from other streams.
After updating each stream to $0'$, $1'$, $2'$, and $3'$, we
reverse the ordering of those, and complete this one update.  We
can see that when we reverse the Markov chain, we would update
$3'$ first with the information from $\F(\{2',1',0'\})$, because
$\F$ does not depend on the ordering of $\X_i$, we can reproduce
the reversed Markov chain.

For the purpose of assisting MD evolution, where $\U$ represents
the procedure of refreshing $p$, integrating the
equation~\eqref{eq:md}, and finishing with a Metropolis-Hastings
accept/reject step, we use
\begin{equation}
  \label{eq:md-assisted-F}
  G^{-1} =
  \F\big(\{\X_i\}_{i=0}^{\n-1}\big) =
  \G^{-1}\left(\sort\big(\{\X_i\}_{i=0}^{\n-1}\big)\right),
\end{equation}
where a special routine $\sort$ sort the set of $\X_i$ before
applying the function $\G^{-1}$, because our choice of $\G$ in
equation~\eqref{eq:md-assisted-G} depends on the ordering of
$\X_i$, and sorting guarantees the same $G^{-1}$ with a reversed
procedure.
In addition to the simple case presented here,
we can build more involved Markov chains by
using more states per parallel stream for the exchange kernel $\F$,
or by decoupling some of the parallel streams for parallel evolution.

\vspace{-0.4em}
\section{\label{sec:bfgs-approx-hessian}L-BFGS approximated Hessian}
\vspace{-0.5em}

Among $\eN$ coupled parallel HMC streams,
for each stream, we use the latest configurations from the other streams
to construct the approximate Hessian with the L-BFGS algorithm.
For these $\eN-1$ configurations,
we compute the site-wise finite differences of lattice gauge fields, $\U_k$,
\begin{equation}
	s_k
		= \log \U_k \U_{k-1}^{-1}
		= \log \U_k \U_{k-1}^\dag, \quad
	y_k
		= \nabla S(\U_k) - \nabla S(\U_{k-1}), \quad
	k=1\ldots L \le\eN-2,
\end{equation}
where $s_k$ and $y_k$ are fields of the elements of the Lie algebra,
$L$ is the length of the L-BFGS memory,
and the inequality comes from
selecting only those field pairs with
$s_k^\T y_k = y_k^\T s_k > 0$
(the inner product implicitly
traces over the color indices and sums over lattice)
for the positive definiteness of the approximated Hessian.
As required in equation~\eqref{eq:md-assisted-F}
we sort the field pairs according to
$s_k^\T y_k$.

The L-BFGS algorithm gives the inverse Hessian as a recursively
defined operator,
\begin{equation}
  \label{eq:bfgs-inv-hessian}
    \G_k^{-1} =
    (I-\rho_ks_ky_k^\T)\G_{k-1}^{-1}(I-\rho_ky_ks_k^\T)
    + \rho_ks_ks_k^\T, \quad
    \G_0^{-1} = 1/(2\beta),
\end{equation}
where $2\beta$ as an initial value comes from the diagonal term
of the Hessian matrix
of the 2-D $U(1)$ action in the
weak coupling limit.  The associated L-BFGS Hessian
matrix can be expressed as,
\begin{equation}
	\label{eq:bfgs-hessian}
	\G_k = \G_{k-1}
		+ \frac{y_ky_k^\T}{y_k^\T s_k}
		- \frac{\G_{k-1}s_ks_k^\T \G_{k-1}}{s_k^\T \G_{k-1}s_k}.
\end{equation}
This rank-2 update has
a symmetric product form~\cite{BRODLIE01021973},
showing the explicit positive definiteness,
\begin{equation}
\label{eq:bfgs-sym-form}
\begin{aligned}
	\G_k&=A_kA_k^\T, &
	\G_0&=A_0A_0^\T, &
	A_k&=(I-\gamma_kv_ks_k^\T)A_{k-1}, &
	\gamma_k&=\rho_k\beta_k, \\
	\G_k^{-1}&=B_kB_k^\T, &
	\G_0^{-1}&=B_0B_0^\T, &
	B_k&=(I-\rho_ks_kv_k^\T)B_{k-1}, &
	\rho_k&=1/(s_k^\T y_k), \\
	\alpha_k&=1/(s_k^\T \G_{k-1}s_k), &
	\beta_k&=\pm\sqrt{\alpha_k/\rho_k}, &
	v_k&=y_k+\beta_k\G_{k-1}s_k.
\end{aligned}
\end{equation}

In our test with the unmodified L-BFGS algorithm, the low eigenvalues of
the approximated Hessian matrix decreases rapidly as the L-BFGS memory
length increases, even after removing all the exact zero modes
from the theory described in the next section.
While the largest eigenvalues are stable, the
condition number increases and the Hessian matrix becomes singular
with a modest L-BFGS memory length,
because the approximate action surface spanned by the
samples we draw for the L-BFGS algorithm has zero modes and may even be
concave.
Since the MD
evolution involves the inverse of the Hessian matrix, the evolution
becomes unstable with near zero modes.
A straightforward method to regulate the approximated Hessian
would be to add a small term to the diagonal of $\G_k$.
It nevertheless breaks the rank-2 update iteration,
invalidates the simple inversion formula and the symmetric
decomposition for the square root of $\G$.
This would require a conjugate gradient inversion
and a rational approximation of the square root of $\G$.

Investigating the determinant behavior from
the symmetric product form~\eqref{eq:bfgs-sym-form}
leads us to one solution: adding a small term to one of the rank-1
updates in equation~\eqref{eq:bfgs-hessian},
\begin{equation}
	\G_k = \G_{k-1}
		+ \frac{y_ky_k^\T}{y_k^\T s_k}
		- \left(
			1 - \lambda\frac{s_k^\T s_k}{s_k^\T \G_{k-1}s_k}
		\right)
		\frac{\G_{k-1}s_ks_k^\T \G_{k-1}}{s_k^\T \G_{k-1}s_k}.
\end{equation}
This still invalidates the iteration formula~\eqref{eq:bfgs-inv-hessian}.
The symmetric product form~\eqref{eq:bfgs-sym-form}, however,
remains applicable with minimal changes.
The complexity of iterating the symmetric product form is
always linear in lattice volume and, in terms of $L$,
$O(L)$ in space and $O(L^2)$ in time.

\vspace{-0.4em}
\section{\label{sec:2du1}$U(1)$ gauge theory on a 2-D lattice}
\vspace{-0.5em}

We use the Wilson plaquette action for the $U(1)$ gauge theory on
a two-dimensional lattice with periodic boundary conditions.
As the QNHMC algorithm uses approximated Hessian, we first need to
remove all the exact zero modes of the Hessian from the theory,
in order to improve the stability of the MD evolution.

The gauge degrees of freedom form the exact zero modes of the Hessian.
With periodic boundary conditions, that is $N_s\times N_t - 1$ zero
modes for a lattice with spatial and temporal extent $N_s$ and $N_t$.
We fix the gauge using a maximal tree of links which we set the
gauge variables $U_{x,\mu}$ to unity.
The maximal tree includes
lattice sites $x = (x_0, x_1)$ and directions $\mu \in \{\hat{0},\hat{1}\}$
satisfying
\begin{equation}
\label{eq:GFix}
\begin{cases}
0 \le x_0 < N_t-1 &
	\text{with $\mu=\hat{0}$ for temporal links,} \\
x_0=0;~ 0 \le x_1 < N_s-1 &
	\text{with $\mu=\hat{1}$ for spatial links.}
\end{cases}
\end{equation}

There are two global gauge degrees of freedom,
due to the abelian nature of the theory,
\begin{equation}
\begin{cases}
U_{x,\hat{0}} \rightarrow U_{x,\hat{0}}\Lambda_0 &
	\text{for } x_0 = N_t-1, \\
U_{x,\hat{1}} \rightarrow U_{x,\hat{1}}\Lambda_1 &
	\text{for } x_1 = N_s-1,
\end{cases}
\end{equation}
where $\Lambda_0$ and $\Lambda_1$ are elements of the $U(1)$ group.
Thus we fix two more gauge links,
$U_{(N_t-1,0),\hat{0}} = U_{(0,N_s-1),\hat{1}} = 1$,
during the MD evolution of QNHMC to remove these two zero modes.

We are interested in observables that are slow to evolve in the Markov
chain, particularly the topological charge.  We use the
definition of the topological
charge~\cite{Flume:1981cw,Panagiotakopoulos:1984qi},
$Q = (\sum_x \Arg P_x)/(2\pi)$,
where the complex argument $\Arg$ takes the principle value of
$(-\pi,\pi)$.  This definition does not apply to exceptional configurations
(with no contribution to the partition function in the continuum
limit) where $P_x=-1$ for some $x$.  On a two dimensional lattice with
periodic boundary conditions, this definition of topological charge
gives exact integer values.

\vspace{-0.4em}
\section{\label{sec:status-plan}Current status, and future plans}
\vspace{-0.5em}

We implement the $U(1)$ gauge theory in the QEX
framework~\cite{Osborn:2017aci}.
We use
PRIMME~\cite{PRIMME} to study the eigenmodes of the exact Hessian matrix
and the L-BFGS approximated one.


The results below come from the 2-D $U(1)$ theory
at $\beta=4.5$ on a lattice of
size $24\times 24$,
with the number of coupled parallel HMC streams, $\eN=10$ and $20$,
using the number of configurations from 8192 to 65536,
depending on the trajectory length.
The MD evolution uses the Omelyan's second order minimum norm
integrator~\cite{OMELYAN2003272}.
We keep the number of steps per MD trajectory fixed
at 8, 16, 32, or 64, while tuning
for optimal trajectory length separately for conventional HMC and QNHMC.

\begin{figure}
  \centering
  \includegraphics[width=0.644\textwidth]{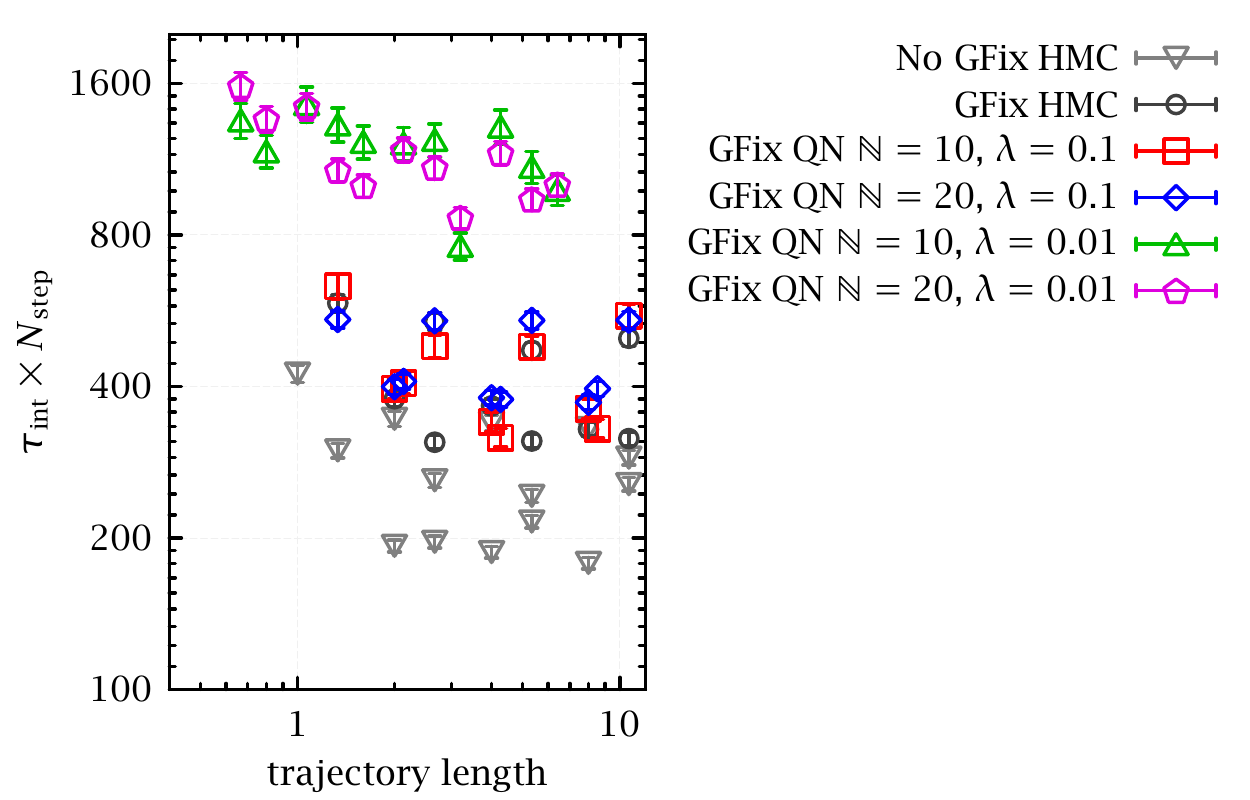}
  \includegraphics[width=0.35\textwidth]{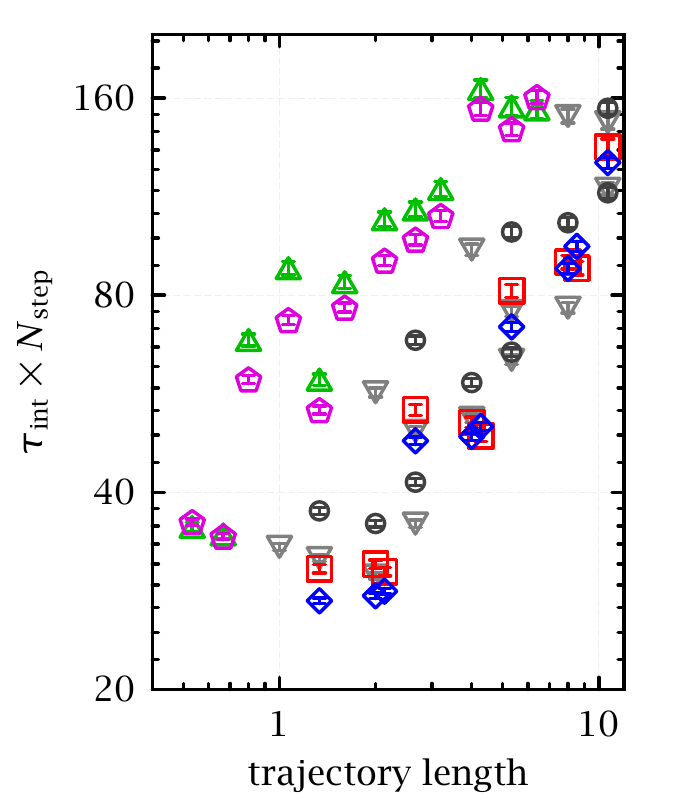}
  \caption{\label{fig:int-autocorr} Integrated autocorrelation length of
	the topologic charge squared (left) and the average plaquette (right) in units of
	discrete MD steps.  GFix denotes gauge fixing during MD.
	QN stands for QNHMC.}
\end{figure}

Figure~\ref{fig:int-autocorr} shows the integrated autocorrelation length
of topologic charge squared (left) and the average plaquette (right).
To include the cost of generating the Markov chain, we multiply the integrated
autocorrelation length by the number of MD steps in an HMC trajectory,
converting the correlation length from the unit of configuration to the unit
of MD steps.
We refer to this quantity as the \emph{cost} of generating 
configurations for uncorrelated measurable quantities,
in terms of the force evaluations,
and we tune simulation parameters to lower the cost.
The apparent increase of the cost
for the average plaquette after the trajectory length
grows longer than three is due to the fact that the autocorrelation becomes
minimal between successive configurations and the cost here
becomes linearly proportional
to the MD steps.

Comparing the conventional HMC with and without gauge fixing, we see
that the topological quantity shows about a factor of two increased cost,
going from no gauge fixing to gauge fixing.
The autocorrelation of the average plaquette value however depends less on
the gauge fixing.
Using QNHMC with $\lambda=0.1$ shows no improvement for
the topological quantity,
and the cost worsens with $\lambda=0.01$.
On the other hand, QNHMC reduces
the cost for the average plaquette with $\lambda=0.1$,
and more so with increased number of coupled Markov chains,
from $\eN=10$ to $20$.

Moving forward, we will do more tuning and testing with the QNHMC
algorithm, studying the scaling behavior toward the continuum limit.
On the other hand, we will also look for other approaches to approximate the
Hessian.  L-BFGS is designed for its efficiency in iterative
optimizations.  Since we have an ensemble of Markov chains, we will
look for other ways to approximate the Hessian
matrix~\cite{2012arXiv1206.6464M,DBLP:journals/corr/MathieuL14}.

\vspace{-0.4em}
\acknowledgments
\vspace{-0.5em}

This research was supported by the Exascale Computing Project
(17-SC-20-SC), a collaborative effort of the U.S. Department of
Energy Office of Science and the National Nuclear Security
Administration.
We gratefully acknowledge the computing resources provided and
operated by the Joint Laboratory for System Evaluation (JLSE) at
Argonne National Laboratory.

\bibliographystyle{JHEP}
\bibliography{r}

\end{document}